\documentclass[doublecol]{epl2} 
\usepackage{graphicx,epsfig}% Include figure files
\usepackage{times}
\usepackage{graphics,dcolumn,bm,float}
\usepackage{amssymb,amsmath,rotate,color}

\title{Reply to ``Comment on Anderson Transition in Disordered Graphene''}
\author{Mohsen Amini\inst{1}, S. A. Jafari\inst{1,2}, Farhad Shahbazi\inst{1}}

\institute{
\inst{1} Department of Physics, Isfahan University of
Technology, Isfahan 84156-83111, Iran\\
\inst{2} The Abdus Salam ICTP, 34100 Trieste, Italy}

\pacs{72.15.Rn}{Localization effects (Anderson or weak localization)}
\pacs{72.20.Ee}{Mobility edges; hopping transport}
\pacs{81.05.Uw} {Carbon, diamond, graphite}

\abstract{}

\begin{document}

\maketitle   

\unitlength 1 cm
\newcommand{\be}{\begin{equation}}
\newcommand{\ee}{\end{equation}}
\newcommand{\bearr}{\begin{eqnarray}}
\newcommand{\eearr}{\end{eqnarray}}
\newcommand{\nn}{\nonumber}
\newcommand{\vk}{\vec k}
\newcommand{\vp}{\vec p}
\newcommand{\vq}{\vec q}
\newcommand{\vkp}{\vec {k'}}
\newcommand{\vpp}{\vec {p'}}
\newcommand{\vqp}{\vec {q'}}
\newcommand{\bk}{{\bf k}}
\newcommand{\bp}{{\bf p}}
\newcommand{\bq}{{\bf q}}
\newcommand{\br}{{\bf r}}
\newcommand{\up}{\uparrow}
\newcommand{\down}{\downarrow}
\newcommand{\fns}{\footnotesize}
\newcommand{\ns}{\normalsize}
\newcommand{\cdag}{c^{\dagger}}

\definecolor{red}{rgb}{1.0,0.0,0.0}
\definecolor{green}{rgb}{0.0,1.0,0.0}
\definecolor{blue}{rgb}{0.0,0.0,1.0}

In a recent comment by Schleede and coworkers~\cite{Comment},
they have correctly pointed out presence of small negative spectral
weights of the order $10^{-4}$ in RKPM expansion of the Dirac
delta function. However, we found that these negative values are not
responsible for the vanishing of the typical density of states (DOS)
near the charge neutrality point. 

\begin{figure}[t]
\includegraphics[angle=-90,width=8.cm,clip=true]{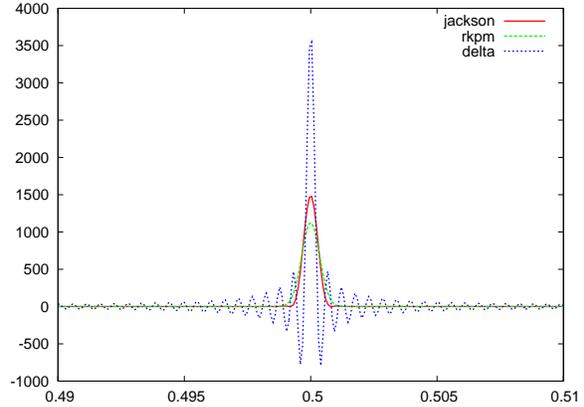}
\caption{Comparison of Dirac delta expansion corresponding to the same expansion order, $N=10^4$.
}.
\label{delta.fig}
\end{figure}

  To clarify this point, in Fig.~\ref{delta.fig} we have plotted
a Dirac delta function with Jackson attenuation factors (Solid line), 
and with RKPM (dashed line). The Gibbs oscillation seen in the dotted line 
are washed out both with Jackson as well as RKPM damping methods. 
As can be seen in the figure, the height of Dirac delta peak 
is underestimated by RKPM relative to the one obtained with Jackson $g$-factor.
Such underestimation of the  LDOSs, along with small $\sigma$ broadening of kernel
might be a possible reason for vanishing of typical DOS in regions with larger
level spacing. Since largest level spacing in the tight-binding spectrum 
of graphene appears around the Dirac point, one expects to obtain vanishingly
small LDOSs giving rise to the mobility edge reported in our letter~\cite{amini}. 
We repeated our calculations with larger kernel width, and we obtain
results similar to those obtained in Fig.~2 of Ref.~\cite{Comment}.
Therefore we agree that the existence of  mobility edge can be due to
subtle details of the kernel used in the expansion. 

%%%%%%%%%%%%%%%%%%%%%%%%%%%%%%%%%%%%%%%%%%%%%%%%%%%%%%%%%%%%%%%%%%%%%%%%%%%%%%%%%%%%%%%%%%%%%%%%%%%%%
%%%%%%%%%%%%%%%%%%%%%%%%%%%%%%%%%%%%%%%%%%%%%%%%%%%%%%%%%%%%%%%%%%%%%%%%%%%%%%%%%%%%%%%%%%%%%%%%%%%%%%

\begin{figure}[t]
\includegraphics[angle=-90,width=8.cm,clip=true]{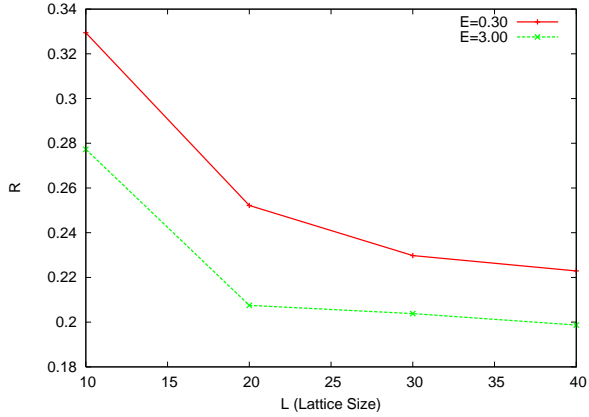}
\caption{Scaling behavior of $R$ for different values of energy with $W/t=12$ in cubic lattice.}
\label{cubic.fig}
\end{figure}

  However, we do not agree with the main conclusion of the comment
that smallest amount of uncorrelated on-site disorder in 2D honeycomb 
lattice is capable of localizing the entire spectrum. To discuss 
this claim, we present the following three arguments: 
(i) Let us repeat the same scaling analysis presented in right
panel of Fig.~ 3 in Ref.~\cite{Comment}. In Fig.~\ref{cubic.fig} we 
show the scaling analysis at fixed energies for $W/t=12$ for a tight-binding
model on a cubic lattice for which the critical value $W_c/t\sim 16.5$.
In agreement with previous works of the above authors~\cite{rmpFehske}, for this value
of disorder, one expects all states to remain extended. 
As can be seen in Fig.~\ref{cubic.fig}, although the ratio 
$R(E)=\rho_{\rm typ}(E)/\rho_{\rm av}(E)$ decreases by increasing system size,
it finally saturates for larger sizes. In absence of a rigorous 
theory for the size dependence  of $R(E)$, the scaling result 
presented in right panel of Fig. 3 of the comment does not conclusively 
imply the localization of all states for the small value of disorder
considered.  

   (ii) The log-normal fitting to the distribution of LDOSs presented
in left panel of Fig.~3 of the comment, shows that peak location of 
the distribution stays more or less fixed at $\ln(\rho/\rho_{\rm av})\approx 0$
by increasing the lattice size. This means that majority of LDOSs are 
equal to the mean value of DOS; a characteristic of extended states.
On the other hand, the width of the 
distribution seems to saturate in the limit of large lattice sizes,
which indicates states with energy $0.25t$ at $W/t=0.3$ are not going
to get localized. 

   (iii) A large body of experimental data support the existence of 
a minimal conductivity in various graphene samples. This is in sharp
contradiction to the main conclusion of the comment. Most spectacular
experimental realization of short range on-site disorder in graphene
was achieved by ARPES analysis of the  Hydrogen dozed graphene 
samples~\cite{RotenbergAL}. The Anderson transition was observed in this 
experiment beyond a certain level of dozing. This experiment
shows that, above a critical disorder strength, state around the Fermi 
level seen in the ARPES are localized as indicated by d.c. conductivity 
measurements~\cite{RotenbergAL}. 

 %%%%%%%%%%%%%%%%%%%%%%%%%%%%%%%%%%%%%%%%%%%%%%%%%%%%%%%%%%%%%%%%%%%%%%%%%%%%%%%%%%%%%%%%%%%%%%%%%%%%%
%%%%%%%%%%%%%%%%%%%%%%%%%%%%%%%%%%%%%%%%%%%%%%%%%%%%%%%%%%%%%%%%%%%%%%%%%%%%%%%%%%%%%%%%%%%%%%%%%%%%%%

   As a closing remark, the question of, whether a mobility edge in 
graphene exists or not~\cite{Naumis,Hilke},
remains open and requires further theoretical and experimental investigations.

%\pagebreak

\end{document}